\documentclass[sigplan,10pt,nonacm]{acmart}

\usepackage{amsmath}
\usepackage{algorithm}
\usepackage{algpseudocode}
\usepackage[table,xcdraw]{xcolor}
\usepackage{booktabs}
\usepackage{array}
\usepackage{stfloats}

\usepackage{colortbl}
\usepackage{array}

\usepackage{multirow}

\algrenewcommand\algorithmicrequire{\textbf{Input:}}
\algrenewcommand\algorithmicensure{\textbf{Output:}}
\algrenewcommand\algorithmiccomment[1]{\hfill$\triangleright$~#1}
\algrenewcommand\alglinenumber[1]{\footnotesize #1:}

\newcommand{\cmt}[1]{%
  \Statex
  \textcolor{gray!55!black}{%
    \makebox[\dimexpr\linewidth-0.5em\relax][l]{%
      \textbf{\texttt{/*  #1}}\hfill\textbf{\texttt{*/}}%
    }%
  }%
}

\AtBeginDocument{%
  }

\begin{document}

\title{ActKV: Efficient LLM Agents through Action-Guided KV Cache Management}


\author{Zihan Wang}
\affiliation{%
  \institution{University of Science and Technology of China}
  \city{Hefei}
  \country{China}}
\email{wangzh196@mail.ustc.edu.cn}

\author{Cheng Tang}
\affiliation{%
  \institution{University of Science and Technology of China}
  \city{Hefei}
  \country{China}}
\email{sisyphustc@mail.ustc.edu.cn}

\author{Lei Gong}
\affiliation{%
  \institution{University of Science and Technology of China}
  \city{Hefei}
  \country{China}}
\email{leigong0203@ustc.edu.cn}

\author{Chao Wang}
\affiliation{%
  \institution{University of Science and Technology of China}
  \city{Hefei}
  \country{China}}
\email{cswang@ustc.edu.cn}

\author{Wenqi Lou}
\affiliation{%
  \institution{Suzhou Institute for Advanced Research, University of Science and Technology of China}
  \city{Suzhou}
  \country{China}}
\email{louwenqi@ustc.edu.cn}

\author{Teng Wang}
\affiliation{%
  \institution{Suzhou Institute for Advanced Research, University of Science and Technology of China}
  \city{Suzhou}
  \country{China}}
\email{wangt635@ustc.edu.cn}

\author{Xuehai Zhou}
\affiliation{%
  \institution{University of Science and Technology of China}
  \city{Hefei}
  \country{China}}
\email{xhzhou@ustc.edu.cn}









\begin{abstract}
Agentic LLM inference accumulates long KV caches across iterative observation-reasoning-action loops, imposing substantial memory overhead and limiting serving throughput. Existing compression methods emphasize overall output quality, overlooking the asymmetric importance of actions in driving task progress. Our key idea is to establish a compression criterion that values KV entries by their contribution to action generation and prioritizes action quality. However, iterative execution, dynamic memory demands, and scattered action-critical entries pose challenges to eviction policies, budget allocation, and paged memory integration. To this end, we propose ActKV, the first KV cache compression framework tailored for agentic LLM inference. (i) Action-oriented KV cache eviction exploits stable action access patterns to retain entries critical to future actions, supporting reliable task progress under compression. (ii) Confidence-driven adaptive budget allocation uses LLM’s intrinsic confidence to adapt the budget to evolving action-critical memory demands. (iii) Page-aware compression management standardizes compression into three primitives with customized kernels, realizing practical throughput gains. On long-trace tasks, ActKV retains an average of 98.53\% of FullKV's accuracy with only 25.98\% of its peak KV cache memory. It also achieves 3.97$\times$ and 3.58$\times$ FullKV's token and task throughput, delivering state-of-the-art performance.
\end{abstract}





\maketitle

\section{Introduction}
\begin{figure}[t]
  \centering
  \includegraphics[width=\linewidth]{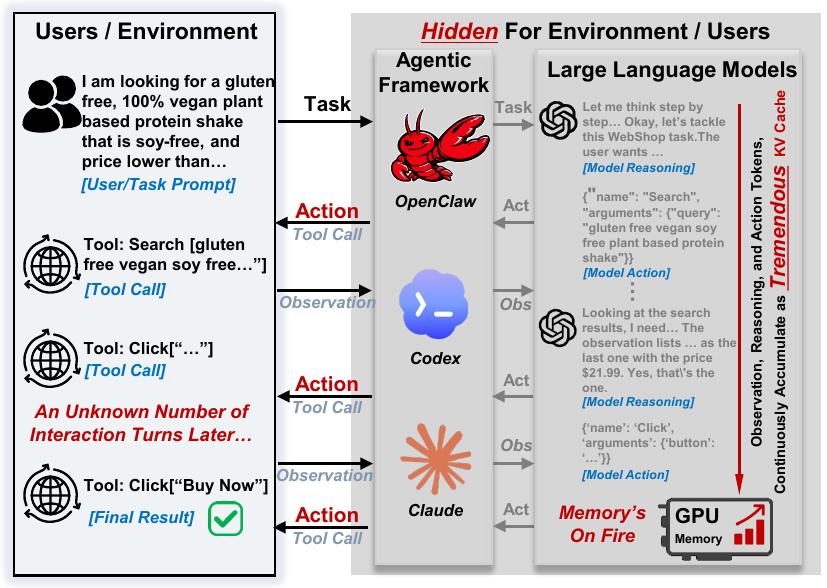}
  \caption{Agentic Workflow}
  \label{fig:intro}
  \Description{...}
\end{figure}

Agents transform LLMs from passive question-answering into task-solving systems that perceive environments, reason, and act, reshaping how users interact with LLMs. The key is to introduce an intermediary agentic layer, forming a three-layer interaction architecture: user/environment, agentic framework, and LLM inference. The fundamental execution paradigm is iterative observation-reasoning-action loops, known as ReAct~\cite{react}, which underpin modern agents like OpenClaw~\cite{openclaw}, Claude Code~\cite{claudecode}, and Codex~\cite{codex}. As shown in Fig.~\ref{fig:intro}, in each iteration, the agentic framework first collects environmental observations and extends LLM prompt. The LLM then reasons via chain-of-thought~\cite{chainofthought} for state understanding and deduction from accumulated traces. Finally, the LLM generates an action, which is executed by the agentic framework in environment. We view agents as complete systems formed by agentic frameworks and LLMs. Agentic LLM inference thus refers to the framework-driven inference process that repeatedly consumes observation tokens and generates reasoning and action tokens. \textit{Importantly, observation and reasoning remain largely internal within the agent, and are often hidden or useless to users, while what users care about is the task progress driven by actions}.

Despite its potential for solving complex tasks, agentic inference incurs substantial KV cache overhead. Observation, reasoning, and action tokens accumulate as KV entries across interaction rounds, imposing heavy memory and computation costs on LLM serving systems. For example, a web shopping task with Qwen3-30B requires 25 iterations and about 35K tokens, consuming 3.20 GB of KV cache, with observation and reasoning entries accounting for over 99\%. This overhead grows with concurrent requests in batched serving, limiting concurrency and throughput. Moreover, long KV sequences can dilute attention and degrade model accuracy. Thus, KV cache compression is essential for efficient and accurate agent deployment.

KV cache compression dynamically evicts redundant KV entries during inference while preserving accuracy. Although many methods exist, most target conventional text generation and generalize poorly to agentic inference. \textbf{The fundamental mismatch lies in the generation objective: preserving overall output quality versus prioritizing action quality}. In continuous text generation, all tokens contribute to output completeness, so existing methods fairly maintain quality of each token generation. By approximating near-future attention, they greedily evict KV entries that are lowly attended by next few tokens~\cite{snapkv, h2o, streamingllm}. \textbf{In agentic inference, however, token importance is asymmetric}. Observation and reasoning tokens mainly support internal state understanding and deduction, but are often hidden or useless for users. \textbf{Only action tokens affect environments and determine user task progress}. Thus, uniform token treatment is inherently misaligned with prioritizing action quality. Even worse, short-sighted retention for near-future reasoning can harm subsequent actions (Section~\ref{sec:Action-Preferred Entries Are Distinct and Stable}).

Our key idea is to establish a \textbf{compression criterion that values each KV entry by its contribution to action generation and prioritizes action quality during compression}, thereby aligning eviction with the outcome-driven nature of agentic inference. However, three challenges arise.

\textit{\textbf{What to keep?} Iterative execution calls for stable criteria to identify action-critical KV entries}. Since successive actions build on accumulated observations and reasoning, compression must preserve KV entries needed for both current and future actions. However, evolving context makes current importance an uncertain indicator of future utility. This raises a fundamental question: do action-critical KV entries exhibit consistent attention patterns across iterations? Such regularities could provide a reliable basis for compression while preserving subsequent action quality.

\textit{\textbf{How much to keep?} Dynamic memory demands of action-critical KV entries require adaptive budget allocation}. The KV cache budget controls the trade-off between memory footprint and generation quality. Across task types and difficulty levels, the number of action-critical KV entries varies with interaction rounds, observation lengths, and reasoning complexity~\cite{react}. However, existing methods typically profile full-KV memory consumption on specific tasks and apply predefined compression ratios during inference~\cite{pyramidkv, adakv}. Such budgets derived from workload-specific offline profiling fail to adapt to evolving action-critical memory demands.

\textit{\textbf{How to compress efficiently?} Scattered action-critical KV entries require reconciling compression with paged memory management}. Agent deployment relies on serving systems (e.g., vLLM~\cite{vllm} and SGLang~\cite{sglang}), which use block-based KV management to reduce fragmentation and improve throughput. However, scattered action-critical entries require token-level eviction misaligned with block-level allocation. Most implementations remain limited to algorithmic validation. The few system-level approaches incur copying overhead offsetting compression gains~\cite{rkv} or rely on policy-specific adaptations lacking generality~\cite{thinkv}. These limitations hinder practical memory savings from fine-grained eviction.

To address the challenges, we propose ActKV, the \textit{first} KV cache compression framework tailored for agentic LLM inference, enabling accurate, memory-efficient, and high-throughput agent deployment through three components.

\textit{Action-oriented KV cache eviction prioritizes action generation quality during compression}. We observe that actions generated within a given task exhibit similarities. First, each task involves a finite set of semantically related action types. Second, action arguments repeatedly draw on shared task context. Third, action outputs follow a standardized template~\cite{toolformat}. Motivated by these semantic, contextual, and structural similarities, we examine attention maps and observe that action-critical KV entries form a stable subset across iterations but contribute intermittently (Section~\ref{sec:Action-Preferred Entries Are Distinct and Stable}). Accordingly, we propose attention-aware Least Recently Frequently Used (LRFU) to retain entries critical to future actions, enabling reliable task progress with a low memory footprint.

\textit{Confidence-driven adaptive budget allocation dynamically adjusts the KV cache budget during inference}. We observe a strong correlation between the LLM's intrinsic confidence and budget sufficiency. Tighter budgets make action-critical KV information incomplete, forcing model to extend reasoning to recover missing context and causing confidence drops (Section~\ref{sec:Confidence Reveals KV Budget Sufficiency}). Accordingly, we design a confidence monitor that combines sliding-window analysis with linear fitting to detect confidence degradation caused by budget insufficiency and trigger budget adjustments, adapting to evolving action-critical memory demands. 

\textit{Page-aware compression management bridges fine-grained KV eviction with block-based memory}. We observe that the eviction-based compression can be distilled into a unified workflow of three primitives: attention calculation, entry eviction, and cache compaction, with calculation and compaction as the main system gaps (Section~\ref{sec:Standardized Page-Aware KV Compression}). Accordingly, we design customized kernels. The recovery-based attention calculation kernel reconstructs attention scores from softmax log-sum-exp (LSE), while the in-place cache compaction kernel performs one-shot in-place copying by conflict-free slot planning. These standardized primitives and customized kernels translate algorithmic gains into tangible memory savings and throughput improvements.

We evaluate ActKV on four agentic LLMs, Qwen3-30B, Qwen3-235B~\cite{qwen3}, GPT-OSS-20B, and GPT-OSS-120B~\cite{gptoss}, across HotpotQA~\cite{hotpotqa}, WebShop~\cite{webshop}, and ALFWorld~\cite{alfworld}. Relative to FullKV, ActKV averages 98.53\% accuracy with only 25.98\% peak KV cache memory on long-trace tasks, compared with at most 74.22\% accuracy and about 33.60\% memory for StreamingLLM~\cite{streamingllm}, SnapKV~\cite{snapkv}, and R-KV~\cite{rkv}. End to end, ActKV improves token and task throughput over FullKV by 3.97$\times$ and 3.58$\times$, respectively.

\begin{itemize}    
\item \textbf{Action-oriented KV cache eviction}. We prioritize action quality during compression to support reliable task progress under aggressive memory constraints.     
\item \textbf{Confidence-driven adaptive budget allocation}. We use LLM's intrinsic confidence to adapt the budget to evolving action-critical memory demands.     
\item \textbf{Page-aware compression management}. We standardize compression into three primitives with customized kernels, realizing practical throughput gains.
\item While maintaining or even improving accuracy, ActKV reduces memory footprint and improves throughput, achieving \textit{state-of-the-art} performance.
\end{itemize}

\section{Background and Motivation}
\subsection{Agentic LLM Inference}
Agents have reshaped human interaction with large language models (LLMs). Unlike conventional question-answering, agents introduce a framework between users/environments and the LLM to repeatedly invoke the model, parse its outputs, and execute actions on behalf of the user. The framework follows iterative observation-reasoning-action loops, known as ReAct~\cite{react}, enabling complex task solving in systems like OpenClaw~\cite{openclaw}, Claude Code~\cite{claudecode}, and Codex~\cite{codex}. Observations provide environmental feedback, reasoning captures internal deduction, and actions provide an explicit interface to affect environments: document search in retrieval-based question answering, button clicks in web tasks, object manipulation in embodied tasks.

Agentic LLM inference thus refers to framework-driven inference that repeatedly consumes observations and generates reasoning and actions. It differs from conventional inference in two key aspects. First, interaction traces accumulating across iterations rapidly expand the KV cache, increasing memory costs while diluting attention and degrading accuracy. Second, token importance is highly asymmetric: observation and reasoning tokens primarily support internal understanding and deduction but are often hidden from or less useful to users, whereas action tokens directly affect environments and determine task progress.

These distinctions respectively highlight the need for KV cache compression and action-oriented optimization in agentic LLM inference.

\subsection{Context Engineering}
Context engineering is an important optimization in agentic frameworks. It reduces the number of tokens entering the model at the text level, thereby mitigating context-window pressure and indirectly slowing KV cache growth. Most context engineering operations preserve the existing trace prefix, so KV cache can still be continuously appended. For example, memory-based methods store important information and retrieve it as additional observations when needed~\cite{mem0, memoryos, amem}. Some operations, such as LLM-based summarization~\cite{summary1, summary2}, rewrite historical context and affect KV cache reuse, but they incur re-prefilling overhead and are typically triggered only occasionally~\cite{prefixcache}, e.g., when the context window approaches its limit. In contrast, eviction-based KV cache compression operates at runtime inference. It does not modify high-level textual context, but reduces storage and computation costs by removing redundant KV entries.

In summary, context engineering and KV cache compression are orthogonal optimizations. Context engineering decides what text enters the model, while KV cache compression decides which runtime KV states are worth preserving.

\subsection{KV Cache Compression}
During autoregressive LLM generation, each new token attends to all previous tokens, whose intermediate states are stored as Key-Value (KV) cache to avoid recomputation~\cite{impress}. As sequences grow, the KV cache expands proportionally, causing substantial memory and computation overhead and making KV cache compression essential~\cite{alisa}. Eviction-based compression removes redundant KV entries, with attention scores serving as the gold-standard importance signal~\cite{h2o,rkv,snapkv}. Formally, for layer $\ell$ and attention head $h$, at decoding step $i$, the current hidden state is projected into query $q_i^{\ell,h}$, while each previous token $s$ has cached key-value pair ($k_{s}^{\ell,h}$, $v_{s}^{\ell,h}$). Let $K_{<i}^{\ell,h}$ $=$ $[$$k_{1}^{\ell,h}$$,$ $\ldots$$,$ $k_{i-1}^{\ell,h}$$]$ and $V_{<i}^{\ell,h}$ $=$ $[$$v_{1}^{\ell,h}$$,$ $\ldots$$,$ $v_{i-1}^{\ell,h}$$]$. The attention score and output are: 
\begin{gather*}
    A_{i}^{\ell,h} = \mathrm{softmax}\left(q_{i}^{\ell,h}(K_{<i}^{\ell,h})^\top\right),\quad
    o_{i}^{\ell,h} = A_{i}^{\ell,h} V_{<i}^{\ell,h},
\end{gather*}
where $A_i^{\ell,h}$ $\in$ $\mathbb{R}^{1 \times (i-1)}$. A larger $A_i^{\ell,h}[s]$ indicates that the cached KV entry of token $s$ contributes more to current token generation, and is therefore more critical for preserving the quality of the current output.

Existing KV cache compression methods generally assume uniform token importance and use near-future attention to greedily evict entries less useful for upcoming tokens. StreamingLLM~\cite{streamingllm} preserves initial and recent tokens, H2O~\cite{h2o} and SnapKV~\cite{snapkv} approximate near-future attention through score accumulation or observation windows. During reasoning, R-KV~\cite{rkv} deduplicates similar entries, whereas CrystalKV~\cite{crystalkv} prioritizes those receiving long-range attention. However, neither readily generalizes to multi-round agentic inference.

Agentic workflows are outcome-driven, where final actions matter more than intermediate reasoning. KV entries useful for near-future reasoning may not support later action generation, making existing methods shortsighted and degrading action correctness and stability under compression.

\subsection{Budget Allocation}
The KV budget controls compressed-cache capacity and defines the trade-off between memory footprint and generation quality. Existing methods typically use workload-specific, offline-profiled static budgets. This is impractical for agentic inference, where the number of action-critical KV entries varies with task type, task difficulty, and model stochasticity. Different environments produce observations of different lengths, difficult tasks require more interaction rounds and longer reasoning, and stochastic actions may lead to divergent future contexts. Thus, a fixed budget may waste memory or over-compress the cache, degrading generation quality.

Some methods study budget allocation, but mainly distribute a predefined total budget across layers or heads. For example, PyramidKV~\cite{pyramidkv} assigns larger budgets to shallow layers with dispersed attention and smaller budgets to deeper layers with focused attention, while Ada-KV~\cite{adakv} extends this idea to head-wise allocation. These methods optimize relative allocation within a fixed budget, but do not determine the overall budget for dynamic agentic workflows.

\begin{figure*}[b]
  \centering
  \includegraphics[width=0.88\textwidth]{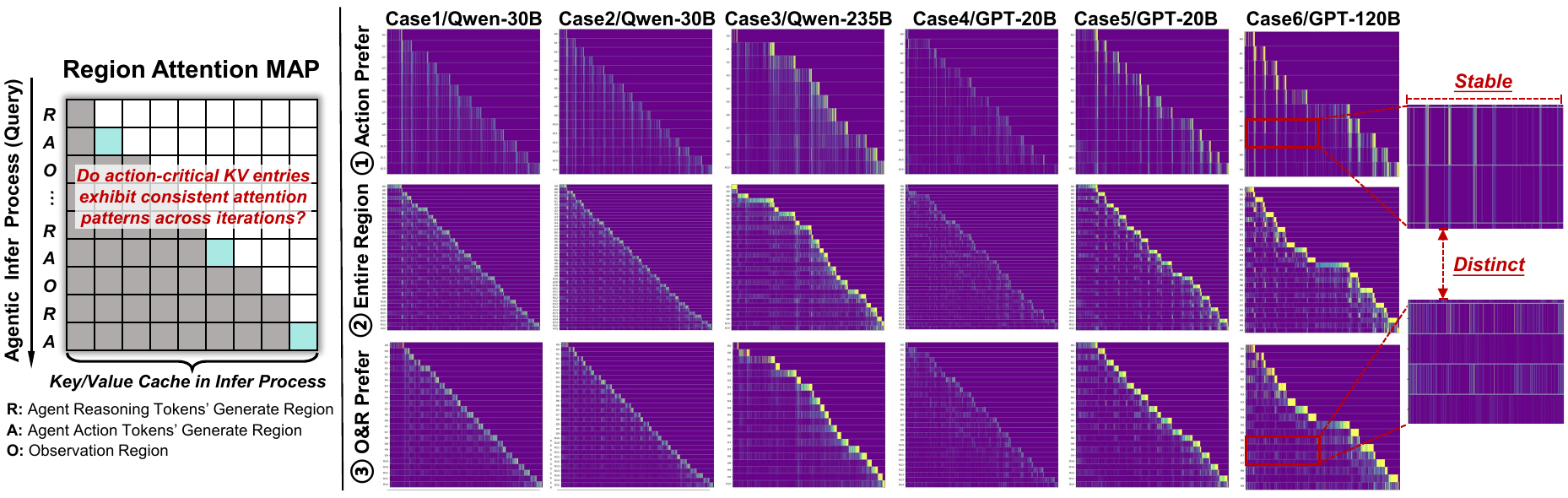}
  \caption{Action-Oriented KV Attention Pattern}
  \label{fig:insightattn}
  \Description{...}
\end{figure*}

\subsection{Paged Memory Management}
Block-based paged KV cache management is common in modern LLM serving systems, with PagedAttention~\cite{pagedattention} as representative design adopted by vLLM~\cite{vllm} and SGLang~\cite{sglang}. It organizes KV cache memory into fixed-size physical blocks:
\begin{gather*}
    \texttt{PagedKVCache[2, B, BS, H, D]}
\end{gather*}
where the first dimension represents key and value states, $\texttt{B}$ is the number of physical blocks, $\texttt{BS}$ is the number of slots per block, $\texttt{H}$ is the number of KV heads, and $\texttt{D}$ is the head dimension. A block table maps logical token positions to physical blocks. For example, when $\texttt{BS}$ $\texttt{=}$ $\texttt{2}$, $\texttt{block\_table}$ $\texttt{=}$ $\texttt{[17,9]}$ and $\texttt{seqused\_k}$ $\texttt{=}$ $\texttt{3}$ correspond to occupied physical slots $\texttt{slot\_map}$ $\texttt{=}$ $\texttt{[34,35,18]}$, leaving one unused slot in last block. By reusing fixed-size blocks, paged KV management avoids per-request contiguous allocation, reducing fragmentation and improving utilization and throughput.

However, action-critical entries are scattered across blocks, making token-level eviction misaligned with block-level allocation. Existing solutions remain limited. R-KV~\cite{rkv} gathers retained entries into an additional contiguous buffer, increasing peak memory usage and offsetting compression benefits. ThinKV~\cite{thinkv} modifies block-table metadata for its compression policy, but algorithm-specific changes disrupt standard serving workflows and limit generality and scalability.

Thus, fine-grained KV compression requires general and scalable integration with paged memory management to translate token-level eviction into actual memory savings and throughput gains.

\begin{figure}[t]
  \centering
  \includegraphics[width=\linewidth]{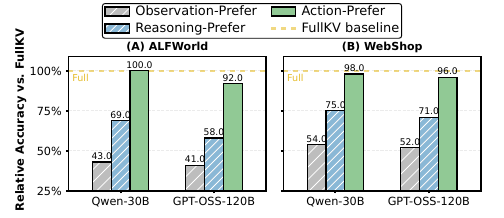}
  \caption{Action-Preferred KV Preservation Study}
  \label{fig:insightevict}
  \Description{...}
\end{figure}

\section{Key Insight}
\subsection{Action-Preferred Entries Are Distinct and Stable}
\label{sec:Action-Preferred Entries Are Distinct and Stable}
We analyze the attention patterns of KV entries that contribute to action generation. As shown in Fig.~\ref{fig:insightattn}, the attention map places key/value states on the horizontal axis and query states on the vertical axis. Query states are partitioned into interleaved observation (O), reasoning (R), and action (A) regions. For each region, we average attention scores along the query dimension to measure each KV entry’s contribution, where brighter colors indicate higher scores and stronger preference. We then extract the top 30\% preferred KV entries for action regions, and apply same procedure to observation and reasoning regions. Across Qwen3-30B, Qwen3-235B, GPT-OSS-20B, and GPT-OSS-120B on ALFWorld and WebShop, we consistently observe two patterns.

First, \textbf{\textit{action-critical KV entries are distinct from those serving observation and reasoning tokens}}. Observation tokens mainly attend to local extended-prefill context. Reasoning tokens aggregate information across all previous observations, producing diffuse attention over history. Action tokens focus on a smaller set of task-critical history for executable decisions. Thus, KV entries important for actions may rank low for observation or reasoning, causing observation or reasoning guided eviction to discard action-critical entries. This explains why normal long-context compression methods fail in agentic inference.

Second, \textbf{\textit{action-critical KV entries are stable across iterations}}. We observe that actions generated within a given task exhibit similarities in their semantics, contextual dependencies, and output structure. Each action consists of three components: type, arguments, and format~\cite{toolcall}. First, each task involves a limited set of semantically related action types, such as clicking in web interaction, object pickup and placement in embodied control, and search and lookup in information retrieval. Second, action arguments repeatedly draw on shared task context, including task constraints, relevant entities, and evidence acquired in previous iterations. Third, action outputs in modern agentic LLMs are often constrained to a common template~\cite{toolformat}. Motivated by these semantic, contextual, and structural similarities, we examine attention maps during action generation and find that action-critical KV entries form a stable subset across iterations (quantified in Section~\ref{sec:Action Access Pattern and Temporal KV Eviction}). This stability makes current action attention an effective signal for preserving KV entries useful for subsequent action generation.

We validate this with a controlled compression experiment. From 100 full-KV-correct traces, we remove the last three iterations’ KV cache and compress the remaining prefix KV to 30\% using observation-, reasoning-, or action-preferred scores. The agent continues for at most six iterations. As shown in Fig.~\ref{fig:insightevict}, action-preferred preservation achieves the best success rate, confirming that action-preferred KV entries are visually distinct, temporally stable, and directly predictive of the entries needed for correct downstream actions.
\begin{figure}[t]
  \centering
  \includegraphics[width=\linewidth]{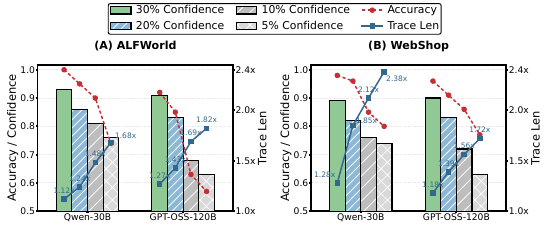}
  \caption{Confidence–Budget Correlation Study}
  \label{fig:insightconf}
  \Description{...}
\end{figure}

\subsection{Confidence Reveals KV Budget Sufficiency}
\label{sec:Confidence Reveals KV Budget Sufficiency}
We analyze the correlation between LLM confidence and KV cache budget. Confidence is derived from the model’s predicted token distribution, where peaked distributions indicate certainty and flatter distributions indicate uncertainty (detailed in Section~\ref{sec:Confidence Monitor}). We take 100 full-KV-correct traces, remove KV cache of last three iterations, and compress remaining prefix KV cache to 30\%, 20\%, 10\%, and 5\% by action preference. We then compare success rate, confidence ratio, and trace-length ratio against full KV. As shown in Fig.~\ref{fig:insightconf}, tighter budgets consistently reduce success rate and confidence while increasing trace length.

We further inspect the generated text traces to understand this behavior. Under tight budgets, compression corrupts action-critical information in KV cache. The model then exhibits low-confidence linguistic markers, such as "wait", "I guess", and "let me check", suggesting that it attempts to extend its reasoning to restate missing context and recover task progress. However, longer reasoning further stresses limited budgets, causing repeated restatement of incomplete history and eventually re-exploration or failure.

This reveals that confidence reflects budget sufficiency in agentic inference. \textbf{\textit{As the budget tightens, action-critical KV information becomes incomplete. The model compensates by extending reasoning to recover missing context, which manifests as confidence degradation}}. Therefore, confidence degradation can serve as a reliable online signal for guiding adaptive budget adjustment during inference.

\subsection{Standardized Page-Aware KV Compression}
\label{sec:Standardized Page-Aware KV Compression}
We distill a unified workflow that integrates token-level KV compression with paged memory management. We observe that most compression methods, including H2O~\cite{h2o}, StreamingLLM~\cite{streamingllm}, SnapKV~\cite{snapkv}, R-KV~\cite{rkv}, RaaS~\cite{raas} and ActKV, can be standardized into three primitives: 
\begin{gather*}
\texttt{AttnScore}\leftarrow \operatorname{CalAttn}(\texttt{targetq},\ \texttt{KCache}), \\
\texttt{KeepMask} \leftarrow \operatorname{Eviction}(\texttt{AttnScore},\ \texttt{Budget}), \\
\texttt{KVCache} \leftarrow \operatorname{Compaction}(\texttt{KVCache},\ \texttt{KeepMask}).
\end{gather*}
Attention calculation computes scores between target queries and cached keys, where different methods use different targets, e.g., observation-window queries in SnapKV and action-specific queries in ActKV. Entry eviction applies a policy over scores and budget to produce a Boolean keep mask, with 1 for retention and 0 for eviction. Cache compaction removes evicted entries and reorganizes remaining KV cache for the subsequent inference.

\begin{figure}[t]
  \centering
  \includegraphics[width=\linewidth]{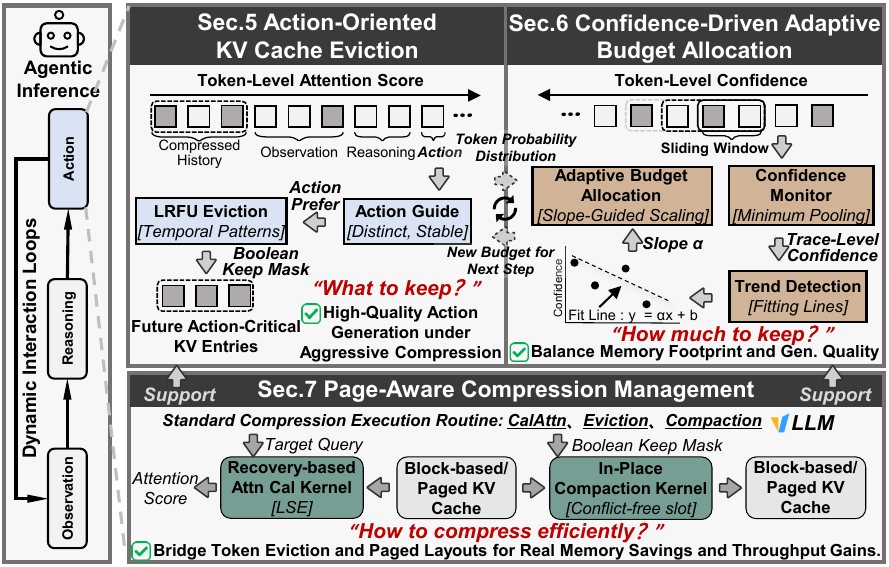}
  \caption{Overview of ActKV}
  \label{fig:overview}
\end{figure}

This decomposition shows that eviction is cache-layout agnostic, while the main system challenges lie in attention calculation and cache compaction. For attention calculation, paged attention backends do not expose full attention scores, whereas standard attention kernels require scattered pages to be materialized into contiguous memory. We address this with a recovery-based attention calculation kernel that reconstructs scores from softmax log-sum-exp (detailed in Section~\ref{sec:Recovery-Based Attention Calculation Kernel}). For cache compaction, arbitrary token eviction creates holes across pages. Gather-based compaction incurs extra memory, while naive in-place compaction suffers from read-write conflicts and cannot be safely parallelized. We address this with conflict-free slot planning and a one-shot in-place compaction kernel (detailed in Section~\ref{sec:In-Place Compaction Kernel}).

Thus, \textbf{\textit{standardizing KV compression into a unified workflow supported by customized kernels provides a general approach to integrating token-level eviction with paged memory management}}, turning fine-grained compression into practical memory savings and throughput gains.

\section{Overview}
Fig. ~\ref{fig:overview} presents ActKV, a KV cache compression framework tailored for agentic LLM inference. ActKV consists of three key modules: action-oriented KV cache eviction, confidence-driven adaptive budget allocation, and page-aware compression management. At the end of each observation, reasoning, action round, the action-oriented eviction compresses KV cache to target budget. Guided by action-region attention scores and attention-aware LRFU policy, it evicts entries that are less important for subsequent action generation, enabling reliable task progress with a low memory footprint (Section~\ref{sec:Action-Oriented KV Cache Eviction}). The confidence-driven adaptive budget allocation continuously monitors model confidence from predicted token distributions via sliding-window scheme. Before each compression step, it detects confidence trend via linear fitting and adjusts budget accordingly, adapting to evolving action-critical memory demands across tasks (Section~\ref{sec:Confidence-Driven Budget Allocation}). The page-aware compression management  bridges compression policies with block-based KV memory layouts via recovery-based attention calculation and in-place compaction kernels, translating fine-grained compression into practical memory savings and throughput gains (Section~\ref{sec:Page-Aware Compression Management}). 

Overall, eviction and budget allocation module form a lightweight feedback loop between generation behavior and KV cache management, while page-aware compression module provides system-level interfaces for efficiently executing compression policies on paged memory layouts. These modules work synergistically to enable accurate, memory-efficient, and high-throughput agent deployment.

\section{Action-Oriented KV Cache Eviction}
\label{sec:Action-Oriented KV Cache Eviction}
\subsection{Action Access Pattern and Temporal KV Eviction}
\label{sec:Action Access Pattern and Temporal KV Eviction}
In Section~\ref{sec:Action-Preferred Entries Are Distinct and Stable}, we qualitatively observed that action-preferred KV entries remain stable across successive action regions. We now quantify this stability with the action access vector ($\texttt{AAV}$) for each KV entry:
\begin{gather*}
    \texttt{AAV}_{kv} = [a_{1},\ a_{2},\ a_{3},\ \dots,\ a_{n}],\ kv \in \texttt{KV\_Cache},
\end{gather*}
where $n$ is the number of action regions that can attend to this entry. Each $a_i$ indicates whether the entry contributes to $i$-th action region: $a_i=1$ if its attention score falls within $\texttt{top-p}$ 90\% of that region, and $a_i=0$ otherwise. We discard all-zero AAVs, as they never contribute to action generation.

\begin{figure}[t]
  \centering
  \includegraphics[width=0.88\linewidth]{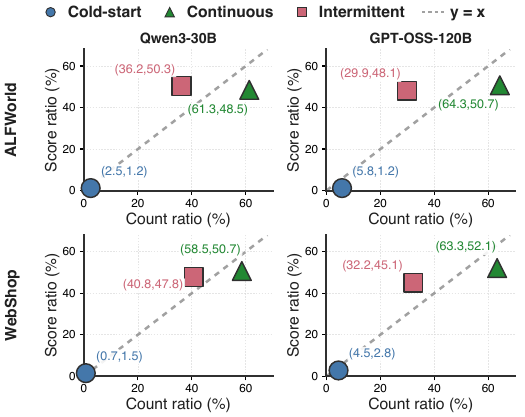}
  \caption{Access Patterns of Action-Preferred KV Entries}
  \label{fig:implevict}
\end{figure}

We classify the remaining AAVs into three mutually exclusive categories. A \textit{cold-start} AAV begins with one or more zeros, indicating that the entry becomes important only in later actions. An \textit{intermittent} AAV contains zeros between its first and last one, indicating discontinuous contribution. A \textit{continuous} AAV contains no zero between its first and last one, indicating consecutive contribution. Fig.~\ref{fig:implevict} reports both the entry-count ratio and per-access attention-score ratio of these categories. Cold-start entries are negligible, suggesting that entries preferred by current actions are usually predictable from previous actions. Continuous entries are the most common, but their per-access attention contribution is lower than that of intermittent entries, which contribute strong attention in a discontinuous manner. Therefore, evicting entries solely based on current action preference may mistakenly remove entries that will become important again.

Motivated by the patterns, we shift from static importance to temporal KV utility. The key criterion is: \textit{a KV entry should be evicted only after it has not contributed high attention for a sufficiently long period}. We instantiate this with attention-aware Least Recently Frequently Used (LRFU) policy. For each KV entry $e_i$, its keep score ($\texttt{KS}$) is updated as:
\begin{gather*}
\texttt{KS}_{t}(i) = \lambda \cdot \texttt{KS}_{t-1}(i) + \texttt{AttnScore}_{t}(i)_{\{\texttt{hit at } t\}},
\end{gather*}
where $t$ is the compression step, $\lambda \in [0,1]$ is the decay factor, and $\texttt{AttnScore}_{t}(i)$ is the action-region attention score of $e_i$. An entry is considered hit if its action-region attention score falls within the $\texttt{top-p}$ set, with $p=90\%$.

This formulation balances recency and frequency. At $\lambda=0$, recency-based eviction may discard temporarily inactive but future-important intermittent entries. At $\lambda=1$, frequency-dominated eviction may retain obsolete continuous entries. With $0<\lambda<1$, high-attention intermittent entries survive temporary inactivity, while score decay enables eviction of stale continuous entries.
\begin{algorithm}[t]
\caption{Action-Oriented KV Cache Eviction}
\label{alg:eviction}
\small
\renewcommand{\baselinestretch}{1.10}\selectfont
\begin{algorithmic}[1]
\Statex\hspace{-\algorithmicindent} \hspace*{-0.5em} \textbf{Inputs:} 
$\texttt{B}$: KV budget, 
$\texttt{p}$: top-$p$ threshold for hit mask, 
$\lambda$: keep-score decay rate, 
$\texttt{Q}^{\texttt{act}}$: action-region queries, 
$\texttt{K}/\texttt{V}$: key/value cache, 
$\texttt{KS}$: keep scores of existing KV entries, 
$\texttt{L}_{ora}$: number of newly generated KV entries in the current ORA round.
\Statex\hspace{-\algorithmicindent} \hspace*{-0.5em} \textbf{Outputs:} 
$\texttt{K}'/\texttt{V}'$: compressed KV cache, 
$\texttt{KS}'$: updated keep scores.
\cmt{Triggered At The End Of}
\cmt{Each Observation-Reason-Action (ORA) Round}
\cmt{Step 1: Get Action-Region Preference}
\State $\texttt{AttnScore} \gets \textsc{PagedCalAttn}(\texttt{Q}^{\texttt{act}},\ \texttt{K})$
\cmt{Step 2: Init Keep Scores for New ORA}
\State $\texttt{KS}_{new} \gets \textsc{Init}(\texttt{L}_{ora},\ 0)$
\State $\texttt{KS} \gets \textsc{Cat}(\texttt{KS},\ \texttt{KS}_{new})$
\cmt{Step 3: Identify Hit and Miss KV Entries}
\State $\texttt{M}_{hit} \gets \textsc{GetHitMask}(\texttt{AttnScore},\ \texttt{p})$
\State $\texttt{M}_{miss} \gets 1 - \texttt{M}_{hit}$
\cmt{Step 4: Update Keep Scores}
\State $\texttt{KS}\texttt{[}\texttt{M}_{hit}\texttt{]} \gets 
\lambda \cdot \texttt{KS}\texttt{[}\texttt{M}_{hit}\texttt{]} + 
\texttt{AttnScore}\texttt{[}\texttt{M}_{hit}\texttt{]}$
\State $\texttt{KS}\texttt{[}\texttt{M}_{miss}\texttt{]} \gets 
\lambda \cdot \texttt{KS}\texttt{[}\texttt{M}_{miss}\texttt{]}$
\cmt{Step 5: Select Retained KV Entries}
\State $\texttt{KeepMask},\ \texttt{KS}' \gets \textsc{TopK}(\texttt{KS},\ k=\texttt{B})$
\cmt{Step 6: KV Cache Compaction}
\State $\texttt{K}'/\texttt{V}' \gets 
\textsc{PagedCompaction}(\texttt{K}/\texttt{V},\ \texttt{KeepMask})$

\State \textbf{return} $\texttt{K}',\ \texttt{V}',\ \texttt{KS}'$
\end{algorithmic}
\end{algorithm}

\subsection{Eviction Algorithm}
Algorithm~\ref{alg:eviction} summarizes the action-oriented eviction procedure. We identify the action region by matching generated tokens to the LLM’s standardized output template to detect the start and end of action generation~\cite{qwen_toolcall, gptoss_toolcall}. We collect attention scores within this region and use them to update the temporal utility of KV entries after each ORA round. Action-critical KV entries differ from those supporting observation and reasoning and remain stable across iterations. Thus, current action attention provides an effective signal for identifying KV entries useful for future actions.

The algorithm updates each KV entry’s keep score using an attention-aware LRFU rule, favoring eviction of entries that have not received high attention for an extended period. This update protects intermittent entries that temporarily become inactive but later receive high per-access attention, while score decay allows stale continuous entries to be evicted. Finally, entries with the highest keep scores are retained, and the KV cache is compacted. This policy exploits stable action access patterns to preserve information critical to future actions under a constrained budget.

\begin{figure}[t]
  \centering
  \includegraphics[width=0.88\linewidth]{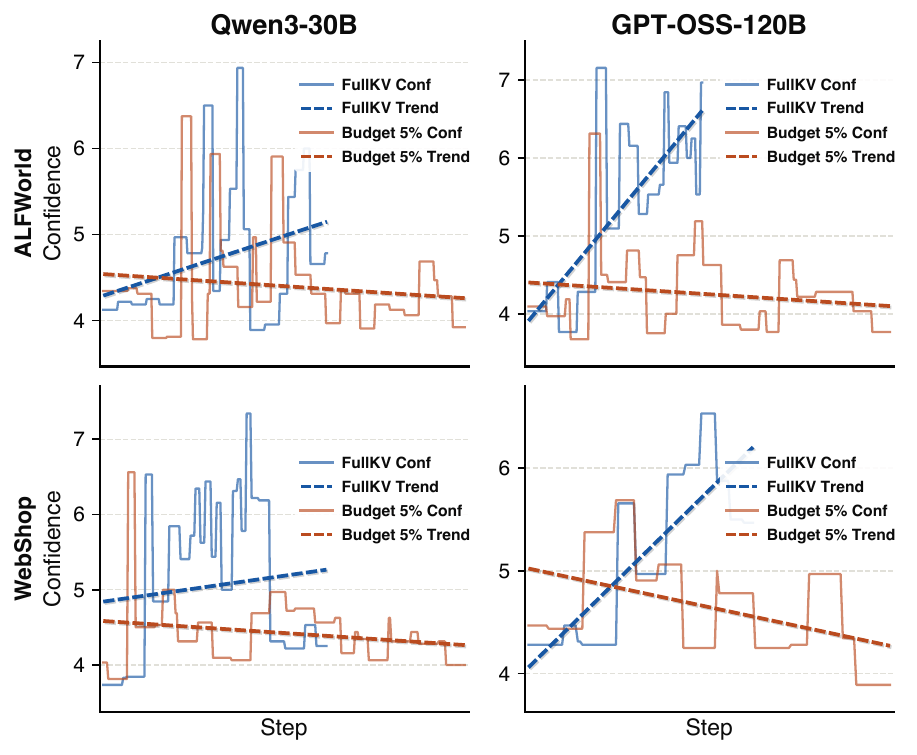}
  \caption{Effectiveness of Confidence Monitor}
  \label{fig:implconf}
\end{figure}

\section{Confidence-Driven Budget Allocation}
\label{sec:Confidence-Driven Budget Allocation}
\subsection{Confidence Monitor}
\label{sec:Confidence Monitor}
Section~\ref{sec:Confidence Reveals KV Budget Sufficiency} shows that insufficient KV budgets corrupt action-critical information, forcing model to extend reasoning to recover missing context and causing confidence degradation. To detect such budget-induced degradation during inference, we introduce a confidence monitor based on internal predicted token distributions~\cite{deepconf}. The monitor operates at three levels: token-level confidence, trace-level confidence, and trend detection across generation steps.

\textit{Token-level confidence}. Given the predicted token distribution $P_{i}$ at position ${i}$, we define token-level confidence as the negative average log-probability of the top-k tokens:
\begin{gather*}
    T_i = -\frac{1}{k}\sum_{j=1}^{k}\log P_i(j),
\end{gather*}
where $P_{i}(j)$ denotes the probability of the $j$-th largest token and $k$ is the number of top tokens considered. High confidence corresponds to peaked distributions and greater model certainty, while low confidence indicates uncertainty in token prediction. 

\textit{Trace-level confidence}. Token-level confidence is local and may fluctuate across tokens, so we aggregate it to characterize the overall confidence of a generated trace. Given a token-level confidence sequence $\{T_i\}_{i=1}^{L}$, we slide a window of size $w$ with stride $s$ and apply minimum pooling:
\begin{gather*}
    R'_{m} = \min_{i \in [ms+1,\dots, ms+w]} T_{i},\ m = 0,1,\ldots,\left\lfloor\frac{L-w}{s}\right\rfloor .
\end{gather*}
This produces an intermediate confidence sequence $R'$. We then retain the bottom b\% values of $R'$ as the final trace-level confidence sequence:
\begin{gather*}
    R = \{ r\in R' | r\leq Q_{b}(R')\},
\end{gather*}
where $Q_b(R')$ denotes the $b$-th percentile of $R'$. Minimum pooling and bottom-$b\%$ selection jointly emphasize low-confidence segments, making the monitor sensitive to localized confidence degradation. Such degradation often occurs when the model restates missing context, repairs its reasoning, or recovers task progress with phrases such as “wait”, “I guess”, or “let me check”.

\textit{Trend detection}. Trace-level confidence can still fluctuate, making absolute values unreliable. We therefore estimate its overall trend by applying first-order least-squares fitting to $R$. Let $\alpha$ denote the fitted slope. If $\alpha<0$, confidence is decreasing, suggesting an insufficient KV budget, so ActKV increases the budget to preserve more entries. The monitor runs in parallel with model inference, introducing negligible overhead to LLM inference.

As shown in Fig.~\ref{fig:implconf}, we apply above monitor to experiments in Section~\ref{sec:Confidence Reveals KV Budget Sufficiency} and compare the confidence trend under a 5\% KV cache budget with that under full-KV cache setting. The monitor accurately captures the confidence degradation caused by insufficient budgets, while the linear fitting effectively reflects the overall trend despite local fluctuations.

\begin{algorithm}[t]
\caption{Confidence-Driven Budget Allocation}
\label{alg:budget}
\small
\renewcommand{\baselinestretch}{1.10}\selectfont
\begin{algorithmic}[1]
\Statex\hspace{-\algorithmicindent} \hspace*{-0.5em} \textbf{Inputs:} 
$\texttt{P}$: predicted distribution, 
$\texttt{k}$: number of top tokens, 
$\texttt{TL}$: token confidence list, 
$\texttt{ws}$: sliding-window size, 
$\texttt{s}$: stride, 
$\texttt{b}$: bottom-$b\%$ ratio, 
$\texttt{B}$: current KV budget, 
$\texttt{scale}_{up}$: budget scaling factor, 
$\texttt{B}_{upper}$: budget upper bound.

\Statex\hspace{-\algorithmicindent} \hspace*{-0.5em} \textbf{Outputs:} 
$\texttt{B}'$: updated KV budget.

\cmt{Monitoring in Parallel with LLM Inference}

\cmt{Step 1: Collect Token-Level Confidence}
\State $\texttt{T} \gets \textsc{GetTokenConf}(\texttt{P},\ \texttt{k})$
\State $\texttt{TL} \gets \textsc{Append}(\texttt{TL},\ \texttt{T})$

\cmt{Step 2: Estimate Trace-Level Confidence}
\State $\texttt{R} \gets \textsc{GetTraceConf}(\texttt{TL},\ \texttt{ws},\ \texttt{s},\ \texttt{b})$

\cmt{Step 3: Detect Confidence Trend}
\State $\alpha \gets \textsc{FitLine}(\texttt{R})$

\cmt{Step 4: Allocate Budget Before Compression}
\State $\texttt{B}' \gets \texttt{B}$
\State \textbf{if} $\texttt{TriggerCompress}$ \textbf{then}
\State \hspace*{0.5em} \textbf{if} $\alpha < 0$ \textbf{then} 
$\texttt{B}' \gets \min(\texttt{B} \cdot \texttt{scale}_{up},\ \texttt{B}_{upper})$

\State \textbf{return} $\texttt{B}'$
\end{algorithmic}
\end{algorithm}

\subsection{Adaptive Budget Allocation Algorithm}
Algorithm~\ref{alg:budget} presents the confidence-driven adaptive budget allocation procedure. It continuously tracks generation confidence in parallel with LLM inference and adjusts the KV cache budget before compression. The algorithm first derives token-level confidence from predicted token distributions and appends it to a running sequence. It then aggregates token-level confidence into trace-level confidence using sliding-window minimum pooling and bottom-percentile selection, emphasizing localized degradation associated with context restatement, reasoning repair, and task-progress recovery. Finally, it fits a linear function to detect the confidence trend and increases the budget when confidence declines, forming a lightweight feedback loop between generation behavior and KV cache management.

As the task progresses, action-critical KV entries accumulate, increasing the memory required to support subsequent actions. We therefore initialize the cache with a small budget and expand it only when confidence declines, accommodating growing memory demands while avoiding premature over-allocation. When confidence stabilizes or recovers, we retain the current budget, as this improvement suggests that the allocation is sufficient but does not establish that a smaller budget would suffice. Maintaining the allocation thus avoids renewed information loss from budget reductions and supports continued confidence recovery.

\begin{algorithm}[t]
\caption{Recovery-Based Attention Calculation Kernel}
\label{alg:calattn}
\small
\renewcommand{\baselinestretch}{1.10}\selectfont
\begin{algorithmic}[1]
\Statex\hspace{-\algorithmicindent} \hspace*{-0.5em} \textbf{Inputs:}
$\texttt{Q}$: current query states,
$\texttt{PagedKV}$: paged KV cache,
$\texttt{BlockTable}$: logical-to-physical block mapping,
$\texttt{LSE}$: softmax log-sum-exp,
$\texttt{GQAGroup}$: query-head group for each KV head,
$\texttt{ActionReq}$: requests in action regions.

\Statex\hspace{-\algorithmicindent} \hspace*{-0.5em} \textbf{Outputs:}
$\texttt{Out}$: running mean attention scores within action region.

\cmt{Triggered For \textsc{PagedCalAttn} Operation}

\cmt{Step 1: Parallel Over ActionReq/Block/Head}
\State \textbf{parallel for} $\texttt{req} \in \texttt{ActionReq},\ \texttt{block},\ \texttt{kv\_head}$ \textbf{do}

\cmt{Step 2: Load Paged KV Block}
\State \hspace*{0.5em} $\texttt{id} \gets \texttt{BlockTable}[\texttt{req},\ \texttt{block}]$
\State \hspace*{0.5em} $\texttt{K}_{block} \gets \texttt{PagedKV}[0,\ \texttt{id},\ \texttt{:},\ \texttt{kv\_head},\ \texttt{:}]$

\cmt{Step 3: Recover for Current Action Query}
\State \hspace*{0.5em} \textbf{for} $\texttt{q\_head} \in \texttt{GQAGroup}(\texttt{kv\_head})$ \textbf{do}
\State \hspace*{1.0em} $\texttt{Q}_{head} \gets \textsc{LoadQuery}(\texttt{Q},\ \texttt{req},\ \texttt{q\_head})$
\State \hspace*{1.0em} $\texttt{Logits} \gets \texttt{Q}_{head} \cdot \texttt{K}_{block}^{\top}$
\State \hspace*{1.0em} $\texttt{Attn} \gets \exp(\texttt{Logits} - \texttt{LSE}[\texttt{req},\ \texttt{q\_head}])$

\cmt{Step 4: Aggregate over Query Heads}
\State \hspace*{1.0em} $\texttt{AttnScore} \gets \max(\texttt{AttnScore},\ \texttt{Attn})$

\cmt{Step 5: Update Action-Region Attention}
\State \hspace*{0.5em} $\textsc{UpdateActionAttn}(\texttt{Out},\ \texttt{AttnScore})$

\State \textbf{return} $\texttt{Out}$
\end{algorithmic}
\end{algorithm}

\section{Page-Aware Compression Management}
\label{sec:Page-Aware Compression Management}
\subsection{Recovery-Based Attention Calculation Kernel}
\label{sec:Recovery-Based Attention Calculation Kernel}
In Section~\ref{sec:Standardized Page-Aware KV Compression}, we formulate token-level KV cache compression as a unified workflow consisting of three primitives: attention calculation, entry eviction, and cache compaction. Attention calculation is a key bottleneck in adapting fine-grained compression to paged memory management. Paged attention backends, such as paged FlashAttention~\cite{pagedflashattn}, support non-contiguous KV blocks but do not expose the full attention matrix needed for eviction. Standard attention kernels provide these scores but require gathering scattered KV blocks into contiguous memory, incurring costly copies and temporary storage.

To avoid recomputing attention from scratch, \textit{our key idea is to recover attention scores from the softmax log-sum-exp (LSE) produced by paged attention}. LSE is a by-product of on-the-fly softmax normalization. For layer $\ell$ and head $h$, at decoding step $i$, given query $q_i^{\ell,h}$ and cached keys $K_{<i}^{\ell,h}$ $=$ $k_{j,\ j<i}^{\ell,h}$, the LSE is:
\begin{gather*}
    \mathrm{LSE}_{i}^{\ell,h} = \log \sum_{j<i} \exp\left(s_{i,j}^{\ell,h}\right), \quad
s_{i,j}^{\ell,h} = q_i^{\ell,h}(k_j^{\ell,h})^\top .
\end{gather*}
where $s_{i,j}^{\ell,h}$ is the unnormalized attention logit. Since attention is obtained by softmax normalization, each attention score can be recovered as:
\begin{gather*}
    A_{i,j}^{\ell,h} = \exp\left(
s_{i,j}^{\ell,h} - \mathrm{LSE}_{i}^{\ell,h}
\right).
\end{gather*}
Thus, during compression, we only recompute local query-key dot products within each paged KV block and normalize them with LSE already produced by paged attention backend. 

Algorithm~\ref{alg:calattn} presents our recovery-based attention calculation kernel under grouped-query attention (GQA). We identify action boundaries by matching generated tokens against the LLM’s standardized output template~\cite{qwen_toolcall, gptoss_toolcall}. The kernel processes only requests currently generating actions and parallelizes over these requests, logical KV blocks, and KV heads. Each thread block handles one KV head of one logical block for one request, resolves its physical block through the block table, and directly loads keys from the paged KV cache. For each associated query head, it computes logits for the current query token and recovers attention probabilities using the corresponding LSE. It then takes the maximum across query heads in the GQA group and incrementally updates the mean across action query tokens, producing a score for each KV entry without buffering past queries. The accumulated state is reset to zero upon entry into each action region. Implemented in Triton~\cite{triton}, this kernel computes token-level attention scores directly on paged KV layouts without gathering scattered blocks into contiguous memory.

\begin{algorithm}[t]
\caption{In-Place KV Cache Compaction Kernel}
\label{alg:compaction}
\small
\renewcommand{\baselinestretch}{1.10}\selectfont
\begin{algorithmic}[1]
\Statex\hspace{-\algorithmicindent} \hspace*{-0.5em} \textbf{Inputs:} 
$\texttt{PagedKV}$: paged KV cache, 
$\texttt{KeepMask}$: keep mask, 
$\texttt{SlotMap}$: logical-to-physical slot mapping, 
$\texttt{SeqUsedK}$: used KV length of reqs, 
$\texttt{B}$: KV budget of reqs, 
$\texttt{BlockSize}$: page block size.

\Statex\hspace{-\algorithmicindent} \hspace*{-0.5em} \textbf{Outputs:} 
$\texttt{PagedKV}$: compacted paged KV cache.

\cmt{Triggered at Action Region Exit}

\cmt{Step 1: Plan Copy Pairs in Parallel}
\State \textbf{parallel for} $\texttt{req} \in \texttt{ActionExitedRequests}$ \textbf{do}
\State \hspace*{0.5em} $\texttt{Y} \gets \left\lceil \texttt{SeqUsedK}\texttt{[}\texttt{req}\texttt{]}\ /\ \texttt{BlockSize} \right\rceil$
\State \hspace*{0.5em} $\texttt{X} \gets \left\lceil \texttt{B}\texttt{[}\texttt{req}\texttt{]}\ /\ \texttt{BlockSize} \right\rceil$
\State \hspace*{0.5em} $\texttt{Start} \gets (\texttt{Y} - \texttt{X}) \cdot \texttt{BlockSize}$
\State \hspace*{0.5em} $\texttt{End} \gets \texttt{Start} + \texttt{B}\texttt{[}\texttt{req}\texttt{]}$

\cmt{Step 2: Identify Logical Destination Slots}
\State \hspace*{0.5em} $\texttt{DstMask} \gets \neg \texttt{KeepMask}\texttt{[}\texttt{req},\ \texttt{Start}:\texttt{End}\texttt{]}$

\cmt{Step 3: Identify Logical Source Slots}
\State \hspace*{0.5em} $\texttt{KeepMask}\texttt{[}\texttt{req},\ \texttt{Start}:\texttt{End}\texttt{]} \gets \texttt{False}$
\State \hspace*{0.5em} $\texttt{SrcMask} \gets \texttt{KeepMask}\texttt{[}\texttt{req}\texttt{]}$

\cmt{Step 4: Map Logical Slots to Physical Slots}
\State \hspace*{0.5em} $\texttt{DstSlots} \gets \texttt{SlotMap}\texttt{[}\texttt{req},\ \texttt{DstMask}\texttt{]}$
\State \hspace*{0.5em} $\texttt{SrcSlots} \gets \texttt{SlotMap}\texttt{[}\texttt{req},\ \texttt{SrcMask}\texttt{]}$

\cmt{Step 5: Generate Conflict-Free Copy Pairs}
\State \hspace*{0.5em} $\texttt{CopyPairs}\texttt{[}\texttt{req}\texttt{]} 
\gets \textsc{Zip}(\texttt{SrcSlots},\ \texttt{DstSlots})$

\cmt{Step 6: Parallel One-Shot In-Place Copy}
\State \textbf{parallel for} $(\texttt{src},\ \texttt{dst}) \in \texttt{CopyPairs}$ \textbf{do}
\State \hspace*{0.5em} $\texttt{PagedKV}\texttt{[}\texttt{:},\ \texttt{dst}\texttt{]} 
\gets \texttt{PagedKV}\texttt{[}\texttt{:},\ \texttt{src}\texttt{]}$

\State \textbf{return} $\texttt{PagedKV}$
\end{algorithmic}
\end{algorithm}

\subsection{In-Place Compaction Kernel}
\label{sec:In-Place Compaction Kernel}
Cache compaction is another key bottleneck in bridging token-level KV compression with paged memory management. Token eviction leaves holes within and across pages, requiring retained KV entries to be packed into contiguous logical slots. Existing gather operators require additional workspace, while naive in-place compaction introduces read-write conflicts that hinder safe parallelization.

\textit{Our key idea is to perform conflict-free slot planning before data movement}. We observe that retained entries are typically denser near the tail of allocated pages after eviction. We therefore compact all retained entries into the last $X$ logical blocks. For a request occupying $Y$ logical blocks before compression, the target region is defined by:
\begin{gather*}
    X = \lceil \texttt{budget}\ /\ \texttt{block\_size} \rceil, \\
start = (Y - X) \cdot \texttt{block\_size}, \\
end = (Y - X) \cdot \texttt{block\_size} + \texttt{budget}.
\end{gather*}
Based on the keep mask, destination slots are evicted positions inside this tail region, while source slots are retained entries outside it:
\begin{gather*}
    dst\_slots = \texttt{keep\_mask}[start:end] == \texttt{False}, \\
\texttt{keep\_mask}[start:end] = \texttt{False}, \\
src\_slots = \texttt{keep\_mask} == \texttt{True}.
\end{gather*}
These logical slots are then translated into physical KV cache slots through the slot map.

\begin{figure*}[t]
  \centering
  \includegraphics[width=\textwidth]{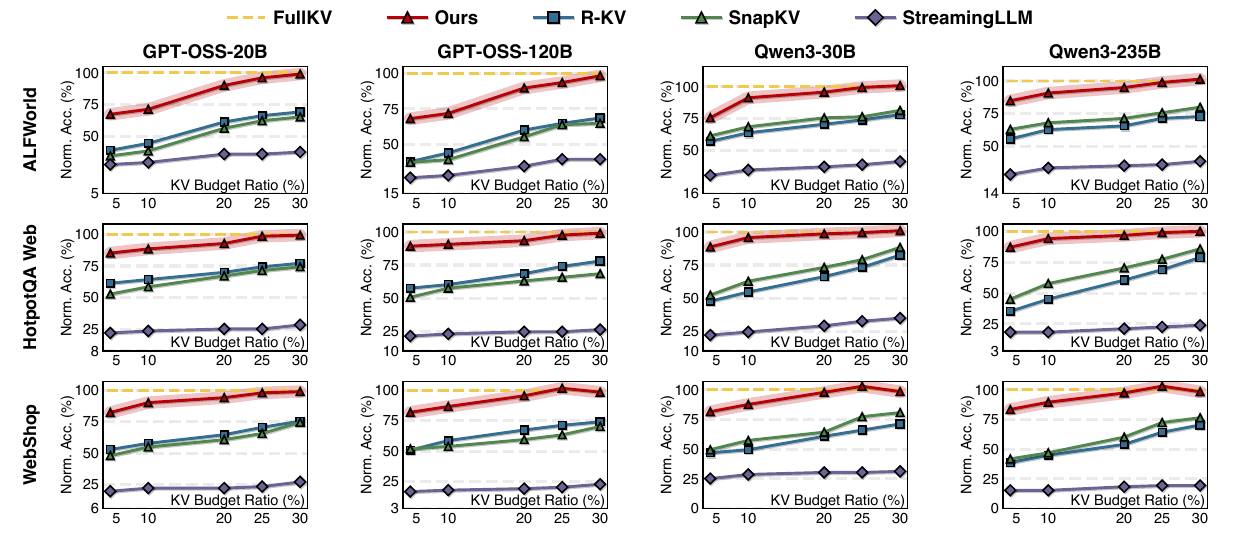}
  \caption{Accuracy Comparison}
  \label{fig:exaccuracy}
  \Description{...}
\end{figure*}

Algorithm~\ref{alg:compaction} presents our in-place KV cache compaction procedure, triggered only for requests that have just exited the action region, after action attention collection and eviction. Slot planning proceeds in parallel across these requests, pairing retained entries outside the target tail region with evicted slots inside it and translating logical slots into physical addresses. A customized copy kernel then performs the planned copies in parallel, with each thread block moving one KV slot. Since source and destination slots are disjoint, all copies complete in one pass without an auxiliary KV buffer. Compaction consolidates retained entries into the tail blocks and returns freed blocks to the serving system for reuse. This memory reclamation is particularly important for agentic inference, where token counts can vary significantly across observation-reasoning-action rounds, leading to fluctuating memory demands. The design efficiently converts token-level eviction into reusable paged memory.

\section{Experimental Evaluation}
\subsection{Experimental Methodology}
\textbf{Models}. We evaluate ActKV on four recent LLMs with strong agentic capabilities: Qwen3-30B-A3B-Thinking-2507-BF16, Qwen3-235B-A22B-Thinking-2507-FP8~\cite{qwen3}, GPT-OSS-20B-BF16, and GPT-OSS-120B-BF16~\cite{gptoss}. These models combine strong reasoning, instruction following, tool use, and structured output, making them representative testbeds for KV compression under agent workloads. We use the official recommended sampling parameters for all models.

\noindent\textbf{Dataset}. We evaluate ActKV on three representative agentic benchmarks: ALFWorld~\cite{alfworld}, HotpotQA-Web~\cite{hotpotqa}, and WebShop~\cite{webshop}. ALFWorld requires sequential actions in a simulated household environment based on textual observations. HotpotQA-Web evaluates multi-hop question answering with external Wikipedia~\cite{wikipedia} retrieval, requiring the model to search, filter, and integrate information across pages. WebShop evaluates web navigation and shopping decisions, where the model compares products and selects an item according to the user goal. For each dataset, we sample 1,500 tasks for evaluation.

\begin{figure*}[t]
  \centering
  \includegraphics[width=\textwidth]{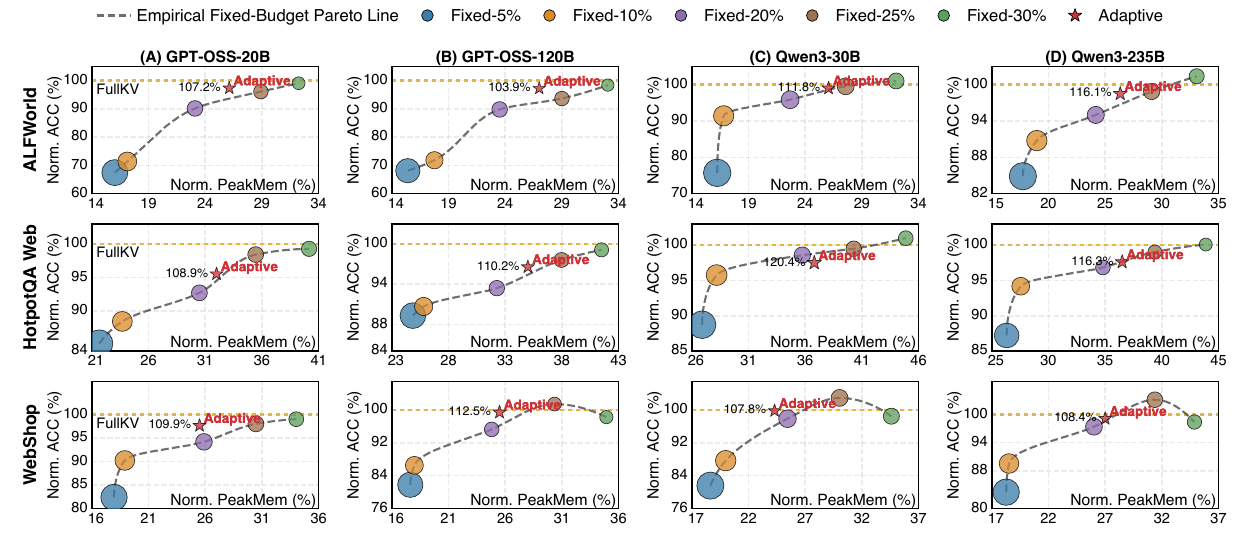}
  \caption{Effectiveness of Adaptive Budget Allocation}
  \label{fig:exbudget}
  \Description{...}
\end{figure*}

\noindent\textbf{Agentic Framework}. We adopt ReAct~\cite{react}, a foundational paradigm for modern LLM agents that interleaves observation, reasoning, and action. For ALFWorld, the action space covers environment observation, navigation, receptacle interaction, object manipulation, and object-state transformation, with an average FullKV trace length of 31K tokens. For HotpotQA-Web, actions include searching Wikipedia entries, locating keywords within pages, and submitting final answers, with an average FullKV trace length of 8K tokens. For WebShop, actions support product search, webpage-link clicking, product-page navigation, and purchasing, with an average FullKV trace length of 33K tokens.

\noindent\textbf{Hardware}. We conduct experiments on a server equipped with eight NVIDIA RTX PRO 6000 Blackwell GPUs (96 GB), Intel Xeon Platinum 8488C CPUs with 192 cores in total, and 256,GB of host memory.

\noindent\textbf{Baselines}. We compare ActKV with four representative baselines: FullKV, R-KV~\cite{rkv}, SnapKV~\cite{snapkv}, StreamingLLM~\cite{streamingllm}. FullKV retains the entire KV cache as an uncompressed reference. SnapKV and StreamingLLM target general long-context generation, while R-KV eliminates redundancy during reasoning. ActKV specifically targets agentic LLM inference. At the implementation level, SnapKV and StreamingLLM primarily provide algorithmic prototypes. R-KV offers a vLLM~\cite{vllm} adaptation, termed paged R-KV, but still gathers KV entries into an auxiliary contiguous workspace outside the paged cache. ActKV supports both Transformers (v5.3.0) and vLLM (v0.19.0), with the latter integrating KV compression directly into paged memory management.

\noindent\textbf{Configuration}. For action-oriented eviction, we set the keep-score decay factor $\lambda=0.5$ and the top-$p$ hit threshold $p=90\%$. For confidence monitoring, we compute token-level confidence from the top-$k$ predicted tokens with $k=20$, then apply sliding-window minimum pooling with window size $w=64$ and stride $s=8$, followed by bottom-$b\%$ selection with $b=10$. The adaptive budget starts at 512 entries and increases by a factor of $\texttt{scale}_{up}=1.75$ when the fitted confidence slope is negative, capped at $B_{upper}=30\text{K}$ entries. We always preserve the KV entries corresponding to the system prompt and user task description, excluding them from compression to retain essential task instructions.

\noindent\textbf{Metrics}. We evaluate five metrics. (1) $\texttt{Accuracy}$ is the fraction of correctly solved tasks, averaged over eight runs per task to mitigate stochastic variability. (2) $\texttt{Budget}$ records the number of KV entries retained after each compression. It remains fixed under static allocation and varies under dynamic allocation. (3) $\texttt{ORA-Len}$ is the number of tokens processed per ORA round. (4) $\texttt{Peak-Memory}$ measures peak KV cache size in entries, given by $\max(\texttt{Budget}+\texttt{ORA-Len})$ across rounds for ActKV, since compression occurs only at ORA boundaries, and $\sum \texttt{ORA-Len}$ for FullKV. (5) $\texttt{Trace-Len}$ is the total number of tokens processed across all rounds of a task, i.e., $\sum \texttt{ORA-Len}$. We report all metrics relative to their corresponding FullKV values.

\subsection{Accuracy Comparison}
Fig.~\ref{fig:exaccuracy} compares the accuracy of ActKV and all baselines, normalized to FullKV. We allocate budgets per trace rather than using dataset-wide average trace lengths, accounting for the substantial variation in agentic trace lengths. Although this offline setting requires prior knowledge of trace length, it enables a controlled comparison of eviction policies. For each trace and budget ratio $r$, we first run FullKV to obtain its length $T$, then run ActKV with budget $B=rT$ and record its mean ORA length $\bar{L}$. ActKV compresses only at ORA boundaries, while each baseline compresses back to the same budget $B$ when its cache exceeds $B+\bar{L}$.

Across all evaluated configurations, ActKV consistently outperforms R-KV, SnapKV, and StreamingLLM. On average, the gains over R-KV, SnapKV, and StreamingLLM are 27.64\%, 27.06\%, and 54.59\% on ALFWorld, 32.86\%, 31.88\%, and 71.08\% on WebShop, and 29.82\%, 28.65\%, and 70.14\% on HotpotQA-Web. On the long-horizon tasks ALFWorld and WebShop, ActKV achieves 81.34\% and 101.41\% of FullKV accuracy with budgets of only 10\% and 25\%, suggesting that action-critical information is concentrated in a small subset of KV entries that ActKV effectively preserves. 

Preserving action-critical KV entries can turn failed trajectories into successful ones. ActKV improves accuracy over FullKV by up to 1.40\% on ALFWorld and 3.13\% on WebShop, indicating that selective retention can improve task success while reducing memory usage. As traces grow, FullKV's expanding cache includes entries less relevant to subsequent actions, potentially diluting attention to action-critical context. These results suggest that ActKV can mitigate this effect by focusing attention on action-critical information, supporting more reliable task progress.

ActKV also generalizes across model families, model scales, and datasets. Its gains are smaller on GPT-OSS than on Qwen3. This difference may partly reflect GPT-OSS's use of sliding-window attention in half its layers, which could increase sensitivity to further KV eviction. ActKV also maintains high accuracy on Qwen3-235B, where each KV head serves more query heads than in Qwen3-30B, supporting its effectiveness at larger model scales.


\begin{table}[t]
\centering
\caption{Effectiveness of Paged Compression Kernels}
\label{tab:excompression}
\renewcommand{\arraystretch}{1.05}
\setlength{\tabcolsep}{6pt}
\begin{tabular}{ccccc}
\toprule
& \textbf{r = 1} & \textbf{r = 4}
& \textbf{r = 16} & \textbf{r = 64} \\
\midrule
1K [0.5K] & 3.18$\times$ & 5.43$\times$ & 25.05$\times$ & 25.50$\times$ \\
2K [1.5K] & 3.76$\times$ & 5.16$\times$ & 15.25$\times$ & 16.47$\times$ \\
6K [4.0K] & 5.63$\times$ & 5.07$\times$ & 7.22$\times$  & 7.18$\times$  \\
\bottomrule
\end{tabular}
\end{table}

\subsection{Adaptive Budget Allocation}
Fig.~\ref{fig:exbudget} compares confidence-driven adaptive and fixed budget allocation in ActKV in terms of accuracy (y-axis), trace length, and peak memory (x-axis), all normalized to FullKV. Per-trace fixed budgets require knowledge of trace lengths from prior FullKV runs and are thus impractical online, whereas adaptive allocation adjusts budgets during inference using confidence feedback. Adaptive ActKV achieves 98.04\%, 96.81\%, and 99.02\% of FullKV accuracy with only 26.38\%, 35.08\%, and 25.58\% of its peak memory on ALFWorld, HotpotQA-Web, and WebShop, respectively. Across the long-horizon tasks ALFWorld and WebShop, it averages 98.53\% of FullKV accuracy with 25.98\% of its peak memory. On both datasets, it lies on the accuracy-memory Pareto frontier~\cite{pareto} among evaluated configurations, demonstrating an effective balance between task success and memory efficiency.

Bubble size denotes trace length, with adaptive ActKV's values relative to FullKV annotated on the left. Tight budgets, such as 5\%, may discard task-critical information, causing persistently low confidence and prolonged reasoning. Since compression occurs only at ORA boundaries, extra reasoning tokens accumulate within each round and may increase peak memory despite a smaller retained budget. ActKV monitors confidence trends and expands the budget upon degradation to mitigate excessive reasoning from insufficient context. Its trace lengths are 109.76\%, 113.97\%, and 109.67\% of FullKV on ALFWorld, HotpotQA-Web, and WebShop, respectively, averaging 109.72\% across ALFWorld and WebShop. Adaptive allocation thus achieves substantial memory savings and near-FullKV accuracy with only modest trace growth.

\subsection{Paged Compression Kernels}
We compare ActKV with paged R-KV to evaluate our customized recovery-based attention calculation and in-place compaction kernels for single compression operations. Regarding memory usage, paged R-KV gathers scattered KV entries into a contiguous workspace, incurring storage and data-copy overhead. ActKV instead operates directly on paged KV layouts without an auxiliary KV buffer. Table~\ref{tab:excompression} reports ActKV's per-operation speedup over paged R-KV. The experiments use $\texttt{block\_size}=16$, $\texttt{kv\_heads}=8$, and $\texttt{head\_dim}=128$, with $r$ denoting the number of concurrent requests, evaluated at 1, 4, 16, and 64. The notation $6\mathrm{K}[4.0\mathrm{K}]$ indicates compressing each request's KV cache from 6K to 4K entries. Across these settings, ActKV achieves an average $10.41\times$ speedup without additional KV storage.

\begin{figure}[t]
  \centering
  \includegraphics[width=\linewidth]{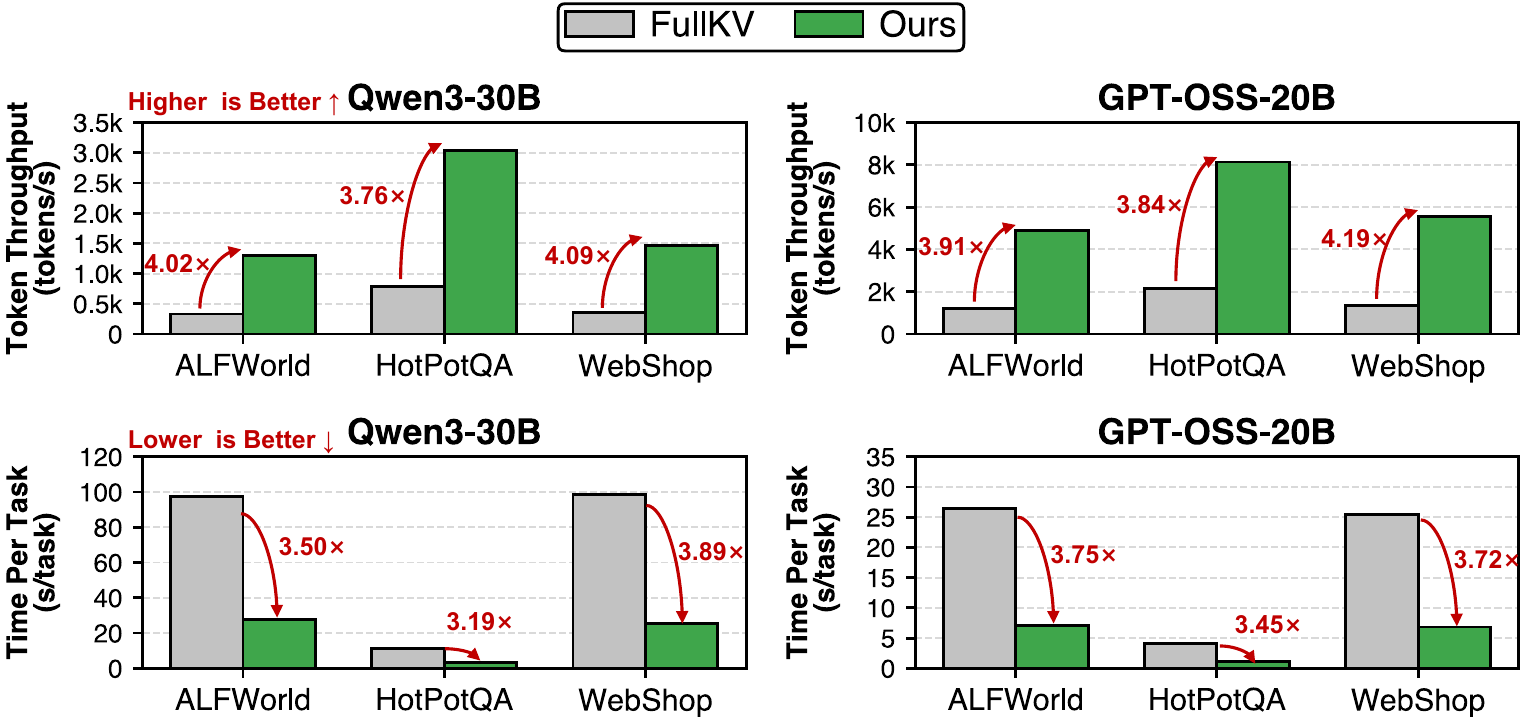}
  \caption{Token and Task Throughput Comparison}
  \label{fig:throughput}
  \Description{...}
\end{figure}

\subsection{End-to-End Throughput}
Fig.~\ref{fig:throughput} compares the offline token and task throughput of ActKV and FullKV using Qwen3-30B and GPT-OSS-20B on a single RTX PRO 6000 Blackwell GPU (96 GB). We submit all 1,500 tasks at once and use the same default vLLM scheduling policy for both methods. ActKV reduces the KV cache footprint through compression and promptly returns freed blocks to the scheduler, allowing memory to be reallocated as requests progress through agentic inference. These mechanisms alleviate memory pressure and enable greater concurrency. Compared with FullKV, ActKV improves token throughput by 3.96$\times$, 3.80$\times$, and 4.14$\times$ on ALFWorld, HotPotQA-Web, and WebShop, respectively. Despite a slight increase in trace length, ActKV also improves task throughput by 3.62$\times$, 3.32$\times$, and 3.80$\times$ on these datasets, respectively.

\section{Conclusion}
This paper presents ActKV, an action-oriented KV cache management framework for agentic LLM inference. ActKV reframes cache management around an agent-specific criterion that values each KV entry by its contribution to action generation, shifting the optimization target toward decisions that drive task progress. Guided by this, ActKV combines action-oriented eviction that exploits stable action access patterns, confidence-driven budget allocation that adapts to evolving memory demands, and page-aware kernels that translate token-level eviction into reusable paged memory. Evaluations across diverse models and agentic benchmarks show ActKV preserves near-FullKV accuracy under aggressive compression while substantially improving serving throughput. More broadly, these findings suggest that efficient agent serving should align memory management with task outcomes rather than token-level fidelity, providing a principled foundation for memory-efficient agentic systems.

\bibliographystyle{ACM-Reference-Format}
\bibliography{ref}










\end{document}